\documentclass[11pt,twoside]{article}
\usepackage{newpasp}

\usepackage{epsf}

\index{Darling, J.}
\index{Giovanelli, R.}

\begin{document}

\title{The Arecibo OH Megamaser Survey and the Galaxy Merger Rate}
\author{Jeremy Darling \& Riccardo Giovanelli}
\affil{Department of Astronomy, Cornell University, Ithaca NY, 14853}
%\author{Riccardo Giovanelli}
%\affil{Department of Astronomy, Cornell University, Ithaca NY, 14853}

% A concise abstract is recommended.  Enter the text of the abstract in
% between the \begin{abstract} and \end{abstract} commands.  Do NOT
% include the word ``Abstract'' in your text; it is inserted
% automatically. Do NOT  make a paragraph break between \begin{abstract} 
% and the first line of the text of the abstract!  Abstracts are required 
% for all papers.

\begin{abstract}
We present the current results of a survey for OH megamasers (OHMs)
underway at the Arecibo Observatory\footnote{The Arecibo
Observatory is part of the National Astronomy and Ionosphere Center, which 
is operated by Cornell University under a cooperative agreement with the
National Science Foundation.}.  The survey is 2/3 complete and has produced
a high OHM detection
rate (1 in 6) from a redshift-selected sample of {\it IRAS} galaxies.
The survey will relate the 
OHM luminosity function to the galaxy merger rate, allowing subsequent
blind OHM surveys to measure the galaxy merger rate as a function of 
cosmic time.  
The survey has also made the first detection of strong variability in OHMs.
Variability will provide a powerful
tool for understanding the small-scale physical settings and mechanisms
of OHMs.
\end{abstract}

% Include keywords if you wish. The keywords.apj file, found on aas.org 
% in the pubs/aastex-misc directory, contains a list of keywords used 
% with the ApJ and Letters.  


% That's it for the front matter.  On to the main body of the paper.

\section{The Survey}
The Arecibo OH megamaser (OHM) survey selects candidates from the PSCz
redshift catalog (Saunders et al. 2000) with the criteria:  
(1) $f_{60\mu m} > 0.6$ Jy, (2) $0.1<z<0.45$, 
and (3) $0^\circ < \delta < 37^\circ$ (Darling \& Giovanelli 2000).  
With a detection rate of 1 OHM
in 6 candidates, the complete survey will double the sample of OHMs to 
roughly 100 objects.  The survey has identified 
35 new OHMs in luminous infrared galaxies to add to the 
sample of 55 found in the literature.  
There is a strong bias for the most FIR-luminous 
galaxies to host OHMs, and a weak FIR color dependence (see Figure
1).  OHMs are detectable out to $z=3$--5 with modern instruments, and
can thus be used to measure the high luminosity tail of the luminous IR galaxy 
luminosity function for redshifts spanning the epoch of major galaxy mergers 
($0.5 < z < 5$).  Blank field surveys for OHMs at various redshifts can 
also measure the galaxy merger rate as a function of cosmic time (Briggs 1998).

Variability has been detected in several OHMs, and is currently
under investigation.  
The variability appears over time scales of months in individual 
spectral features rather than in broad-band modulation which could be 
attributed to antenna calibration or pointing errors.  Variability in 
OHMs constrains the sizes of the variable and quiescent spectral
features, regardless of the source of modulation (intrinsic to the source
or due to propagation effects).  Intrinsically variable regions would 
have sufficiently small angular sizes that they would also be expected
to scintillate (see Walker 1998).   
We thus attribute the variability to interstellar
scintillation, which gives a weaker constraint on the sizes of emission
regions than intrinsic variability.  Variability
in OHMs, particularly those with $0.1<z<1.0$, will provide a powerful
tool for understanding the small-scale physical settings and mechanisms
of masers which can be observed at cosmological distances. 
% For examples on including figures, see the file vla2000_sample.ps
% at http://www.nrao.edu/vla2000/proceedings/. 
% For examples of figures, equations or tables, please see the file
% vla2000_man.ps at the same site. Also available as
% newpaspman.ps at http://www.aspsky.org/pubs/authors.html

% comment this out if you want to include acknowledgements
\acknowledgements
{\small 
The authors are very grateful to Will Saunders for access to the PSCz catalog
and to the excellent staff of NAIC for observing assistance and support.  
This research was supported by Space Science Institute archival grant 
8373 and made use of the NASA/IPAC Extragalactic Database (NED) 
which is operated by the Jet Propulsion Laboratory, California
Institute of Technology, under contract with the National Aeronautics 
and Space Administration.  }

\begin{figure}[t!]
\plotone{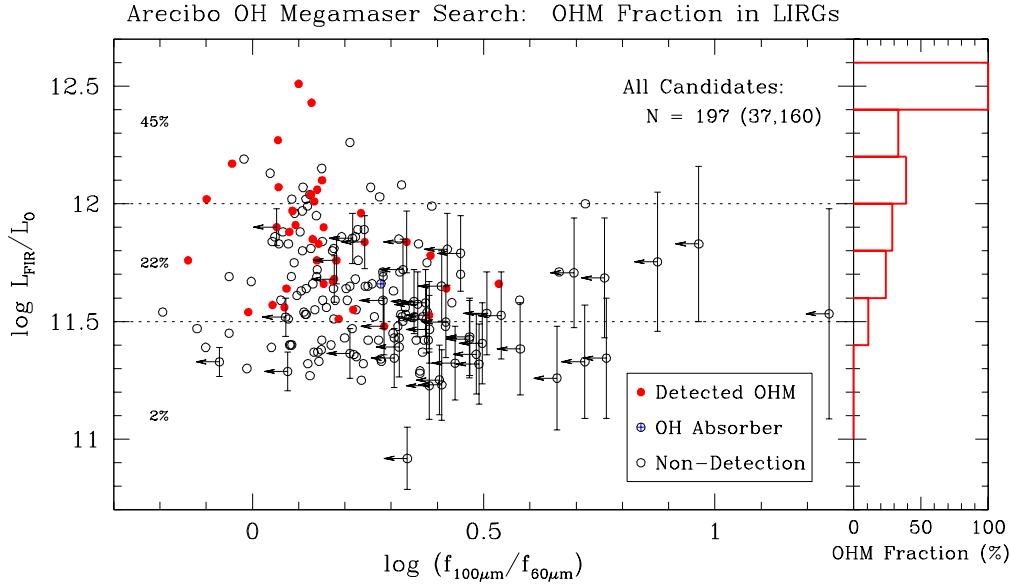}
\caption{\small
Observed OH Megamaser Candidates. The left panel shows $L_{FIR}$ versus 
FIR color for 197 candidates observed to date, and the right panel shows 
the OHM fraction as a function of $L_{FIR}$. 
Points with error bars are non-detections at 100$\mu$m. Vertical error bars 
indicate the possible range of $L_{FIR}$, constrained by $f_{60\mu m}$ 
and an upper 
limit on $f_{100\mu m}$. Horizontal arrows indicate upper limits on FIR color. 
Inset percentages indicate the OHM fraction for each sector delineated 
by the dashed lines. The inset numbers follow
the key: N = Observed (OHMs, Non-Detections).}
\end{figure}

\begin{references}
{\small
\reference Briggs, F.  1998, \aap, 336, 815
\reference Darling, J. \& Giovanelli, R.  2000, \aj, 119, 3003
\reference Saunders, W., et al.  2000, in Cosmic Flows: Towards an Understanding of the Large-Scale Structure in the Universe, ed. S. Courteau, M. Strauss \& J. Willick (San Francisco: ASP), in press
\reference Walker, M. A.  1998, \mnras, 294, 307
}
\end{references}
\end{document}